\RequirePackage[overload]{textcase} 
\RequirePackage[bookmarksnumbered,unicode]{hyperref}
\documentclass[sigconf,letterpaper,natbib=false]{acmart}
\usepackage{wasysym}
\usepackage{fontawesome}
\usepackage{utfsym}
\usepackage{pdfpages}
\usepackage{xspace}
\usepackage{adjustbox}
\usepackage{tcolorbox}
\usepackage{enumitem}
\usepackage[overload]{textcase}

\newcommand{\newcheckmark}{\usym{1F5F8}}
\newcommand{\newcrossmark}{\scalebox{0.75}{\usym{2613}}}

\newcommand{\parvspace}{\vspace{0.20cm}}

\def\name{\textsc{SIMulator}\xspace}

\AtBeginDocument{%
  \providecommand\BibTeX{{%
    \normalfont B\kern-0.5em{\scshape i\kern-0.25em b}\kern-0.8em\TeX}}}

\copyrightyear{2025}
\acmYear{2025}
\setcopyright{rightsretained}
\acmConference[WiSec 2025]{18th ACM Conference on Security and Privacy in Wireless and Mobile Networks}{June 30-July 3, 2025}{Arlington, VA, USA}
\acmBooktitle{18th ACM Conference on Security and Privacy in Wireless and Mobile Networks (WiSec 2025), June 30-July 3, 2025, Arlington, VA, USA}
\acmDOI{10.1145/3734477.3736151}
\acmISBN{979-8-4007-1530-3/2025/06}

\RequirePackage[
  datamodel=acmdatamodel,
  style=acmnumeric,
  ]{biblatex}
  
\begin{document}

\title{\name: SIM Tracing on a (Pico-)Budget}

\author{Gabriel K. Gegenhuber}
\affiliation{%
  \institution{University of Vienna}
  \department{Faculty of Computer Science}
  \department{Doctoral School Computer Science}
  \city{Vienna}
  \country{Austria}
}

\author{Philipp É. Frenzel}
\affiliation{%
  \institution{SBA Research}
  \city{Vienna}
  \country{Austria}
}

\author{Adrian Dabrowski}
{%
\affiliation{%
  \institution{University of Applied Sciences \\ FH Campus Wien}
  \city{Vienna}
  \country{Austria}
}

\renewcommand{\shortauthors}{Gabriel K. Gegenhuber, Philipp É. Frenzel, and Adrian Dabrowski}

\begin{abstract}

SIM tracing --the ability to inspect, modify, and relay communication between a SIM card and modem-- has become a significant technique in cellular network research.
It enables essential security- and development-related applications such as fuzzing communication interfaces, extracting session keys, monitoring hidden SIM activity (e.g., proactive SIM commands or over-the-air updates), and facilitating scalable, distributed measurement platforms through SIM reuse.
Traditionally, achieving these capabilities has relied on specialized hardware, which can pose financial and logistical burdens for researchers, particularly those new to the field.

In this work, we show that full SIM tracing functionality can be achieved using only simple, widely available components, such as UART interfaces and GPIO ports.
We port these capabilities to low-cost microcontrollers, exemplified by the Raspberry Pi Pico (4~USD). Unlike other approaches, it dramatically reduces hardware complexity by electrically decoupling the SIM and the modem and only transferring on APDU level.

By significantly reducing hardware requirements and associated costs, we aim to make SIM tracing techniques accessible to a broader community of researchers and hobbyists, fostering wider exploration and experimentation in cellular network research.

\end{abstract}



\maketitle

\setlength{\textfloatsep}{0.7em plus 1em}
\setlength{\parskip}{0pt plus 3pt}

\section{Introduction}
The capability of inspecting, rewriting, and relaying SIM card communication, also known as SIM tracing, has proven to be a vital tool for cellular network researchers.
Among others, it has been used for
i) fuzzing communication interfaces~\cite{lisowski2024simurai},
ii) extracting session keys or observing hidden SIM card communication~\cite{chalakkal2017practical}, and
iii) reusing SIM cards at different locations (via SIM tunneling) to build scalable measurement platforms~\cite{gegenhuber2023mobileatlas, gegenhuber2022zero}.

Although professional equipment is already available for these operational scenarios (e.g., the Osmocom SIMtrace~2\footnote{\url{https://osmocom.org/projects/simtrace2/wiki}}), not all research labs have access to such specialized hardware, thereby introducing an entry barrier for aspiring researchers or hobbyists entering the field of cellular network research.

In contrast to existing projects, our approach reduces both complexity and cost by making the two ends of the SIM tunnel electrically independent --decoupling voltage levels, communication speeds, and clock domains. Communication is handled exclusively at the APDU level, enabling a clean and modular interface.

By introducing the
MobileAtlas measurement framework~\cite{gegenhuber2023mobileatlas}
for distributed large-scale measurements within mobile networks, we have done a first step in proving that only basic hardware (i.e., a UART interface and GPIO ports) is enough to accomplish the same SIM tracing capabilities.
While MobileAtlas provides an all-in-one experimentation solution based on the Raspberry Pi 4, many applications only require a portion of its functionality, specifically the SIM tracing and relaying capabilities.
In an attempt to popularize SIM tracing and lower the entry barrier, we spun of the concept from MobileAtlas and reimplemented it on super low-cost equipment,
(i.e., the 4~USD Raspberry Pi Pico), encouraging other researchers to take a dive into this topic.

\section{\textsc{\NoCaseChange{{SIMulator}}}}

The primary goal of \name is to lower the cost and entry barrier for SIM tracing by utilizing readily available and inexpensive hardware.
Building on the existing MobileAtlas codebase\footnote{\url{https://github.com/sbaresearch/mobile-atlas}}, \name reduces the tight coupling between the SIM tunnel and the measurement platform, enabling greater modularity and flexibility.
To further broaden potential use cases, we extend support beyond SIM cards to other types of contact smartcards (i.e., both T=0 and T=1 cards), thereby also enabling the introspection and relaying of payment card communication (cf. Figure~\ref{fig:generic-smartcard-emulation} in the Appendix).

\subsection{Architecture}
The \name{}'s architecture (Figure~\ref{fig:simulator-architecture}) is structured as follows:
\begin{itemize}[leftmargin=1em,itemsep=0pt plus 3pt]
    \item A \textbf{Modem} (or smartphone) for which the SIM card is emulated.
    \item A \textbf{Raspberry Pi Pico} connected to the modem's SIM slot to expose the corresponding SIM interface over USB.
    \item A \textbf{Relaying Script} running on a host system to forward communication between the USB interface and  
    \item the MobileAtlas-based \textbf{SIM Provider} that finally terminates the SIM communication via a connected SIM card or eSIM.
\end{itemize}

\noindent This low-cost architecture enables full SIM tracing capabilities using inexpensive, readily available hardware.

\parvspace
\noindent
\textbf{Decoupling Architecture.}
The key to keeping costs low is our system’s reduced complexity compared to other projects (see Table \ref{tab:tool-comparison}). SIM cards operate with various voltages, clock speeds, and dividers, which complicates implementation. Our architecture simplifies this by electrically separating the SIM reader (connected to the SIM) from the SIM simulator (connected to the modem), allowing them to operate and negotiate parameters independently, with only APDU commands exchanged. They can also be hosted on different machines linked via TCP.

\parvspace
\noindent
\textbf{Synchronous (USART) and Asynchronous (UART) Modes.}
The Raspberry Pi Pico can operate in two modes when connecting to the modem: i) a synchronous mode, which requires an additional wire but automatically adapts to different clock (CLK) frequencies, and ii) an asynchronous mode, which simplifies wiring but requires one-time manually setting the CLK frequency (based on a previous frequency measurement e.g., with an Oscilloscope, as done in \cite{gegenhuber2023mobileatlas}).

\begin{figure}[t]
    \centering
    \includegraphics[width=0.9\linewidth,clip=true,trim=0mm 2mm 0mm 2mm]{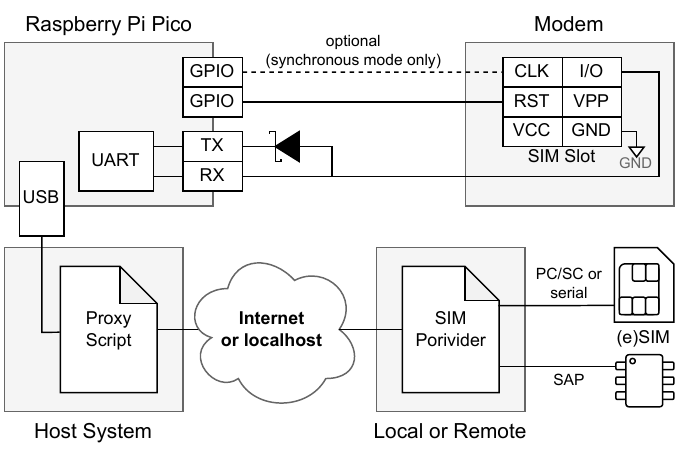}
    \caption{Wiring diagram between our microcontroller and the modem.
    The U(S)ART I/O pins are combined with a Schottky diode to create an open-collector bidirectional bus.}
    \label{fig:simulator-architecture}
\end{figure}

\parvspace
\noindent
\textbf{Full SIM Tracing Capabilities.}
As shown in Table~\ref{tab:tool-comparison}, \name supports tracing (i.e., inspecting), rewriting, and relaying the APDU commands that are sent between the SIM card and the modem.
Compared to existing solutions, it offers improved flexibility at a significantly lower cost.

\parvspace
\noindent
\textbf{Enabling eSIMs on \name.}
In addition to supporting physical SIM cards (e.g., via PC/SC- or super-cheap serial-based card readers), we implemented support for the SIM Access Profile (SAP)~\cite{bluetooth_sim_access_2003}, a legacy protocol used in older cars, available on selected Android smartphones (e.g., Google Pixel devices).
SAP allows our SIM Provider to request direct (APDU-level) access to one of the smartphone's SIM cards, including eSIMs.

By keeping the eSIM in vitro on the smartphone, one can also examine app-controlled eSIMs. For example, such apps might replace the IMSI or the entire eSIM with a new (domestic) one at each border transit, thereby eliminating roaming (e.g., Google Fi in major markets). In contrast, workarounds using physical SIM cards with an embedded eSIM IC cannot be updated on the fly once they are removed from the original phone.

\subsection{Evaluation}
We successfully evaluated \name using seven different modems (Quectel RM520N-GL, Quectel RM500Q, Quectel EG25-G, Huawei ME909s-120, Telit LE910, SIMCom SIM7600E-H, and Sierra MC8355) as well as four distinct smart card terminals.
In all cases, APDU inspection, rewriting, and relaying functions worked reliably in both synchronous and asynchronous modes.

SIM tunneling inherently introduces round-trip latency. %
We implemented ISO~7816 Waiting Time eXtensions (WTX) to deal with those high-latency conditions.
We tested this setup across multiple operators with an artificial delay of 1,000\,ms, and observed no failures or degraded behavior. Thus, the system is also fit to tunnel SIM card communication over satellite Internet connections.

\begin{table}[]
    \centering
    \definecolor{highlight}{RGB}{255,229,229}
    \newcommand\tH{\cellcolor{highlight}}
    \newcommand\tYes{\newcheckmark}
    \newcommand\tNo{\newcrossmark \tH}
    \newcommand\tFa{$^{\ast}$}
    \newcommand\tFb{$^{\dag}$}
    \begin{adjustbox}{width=0.95\columnwidth}
    \begin{tabular}{@{}lccc@{}}
    \toprule
                  & SIMtrace~2          & SIM Interposer\tFa         &  \textsc{SIMulator}            \\ \midrule
    Approx. Price       & 120 USD & 4 USD & 4 USD \\
    APDU Inspection     & \tYes   & \tNo  & \tYes \\
    APDU Rewriting      & \tYes   & \tYes & \tYes \\
    APDU Relaying       & \tYes   & \tNo  & \tYes \\
    Android eSIM        & \tNo    & \tNo  & \tYes \\ 
    Galvanic Separation & (\tYes)\rlap{\tFb}   & \tNo  & \tYes \\
    \bottomrule
    \end{tabular}
    \end{adjustbox}
    \vspace{1ex}
    \begin{minipage}[l]{\textwidth}\footnotesize
    \tFa~Used by \cite{lisowski2024simurai}, e.g. \url{https://turbosim.cn/collections/heicard}
    \hspace{1ex}
    \tFb~Possible with two units.\\
    \end{minipage}
    \caption{Capability comparison of SIM introspection and relaying tools. All costs for the device only, excluding cables.
    }
    \label{tab:tool-comparison}
\end{table}

\section{Conclusion}

In this work, we presented \name, a lightweight and cost-effective platform for SIM tracing that significantly lowers the entry barrier for researchers.
By relying exclusively on readily available components costing just 4 USD, we make advanced SIM tracing capabilities accessible to a much broader audience.
The open-source nature of the project\footnote{\url{https://github.com/sbaresearch/mobile-atlas\#simulator}} further encourages reuse, adaptation, and extension by the research community.
Looking ahead, we envision extending \name toward full SIM virtualization (e.g., as done in \cite{lisowski2024simurai}), enabling complete emulation of SIM functionality without requiring a physical SIM card. %

\printbibliography

\section*{Acknowledgments}
We want to thank Fabian Funder for his practical work on \name.

SBA Research (SBA-K1 NGC) is a COMET Center within the COMET – Competence Centers for Excellent Technologies Programme and funded by BMIMI, BMWET, and the federal state of Vienna. The COMET Programme is managed by FFG.

\appendix

\section{Appendix}
\label{sec:appendix}

\subsection{Supported SIM Reader Devices}
\begin{figure}[h]%
    \centering
	\includegraphics[width=.7\linewidth]{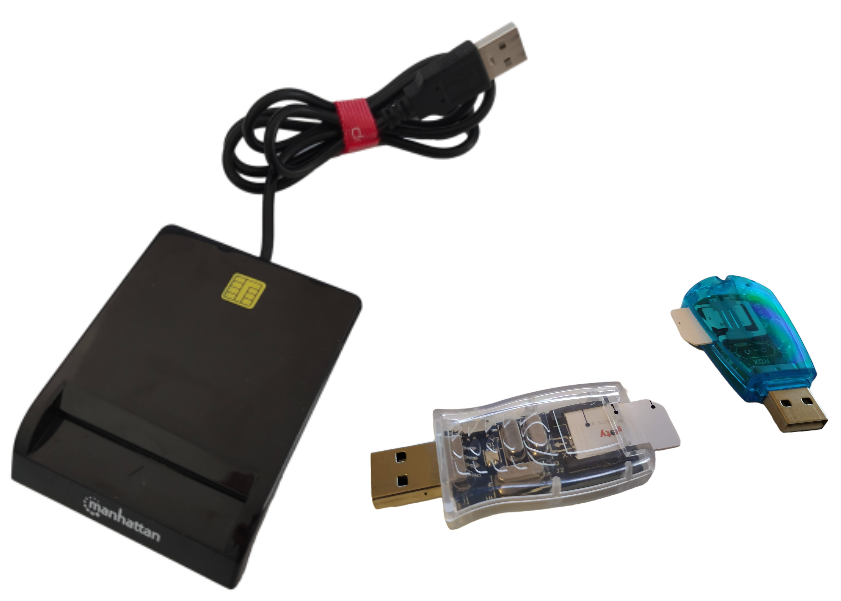}
    \caption{Low-cost SIM reader devices (PC/SC- or serial-based) that are supported by the MobileAtlas-based SIM-Provider.}
    \label{fig:various-sim-reader}
\end{figure}

\begin{figure}[h]%
    \centering
	\includegraphics[width=.7\linewidth]{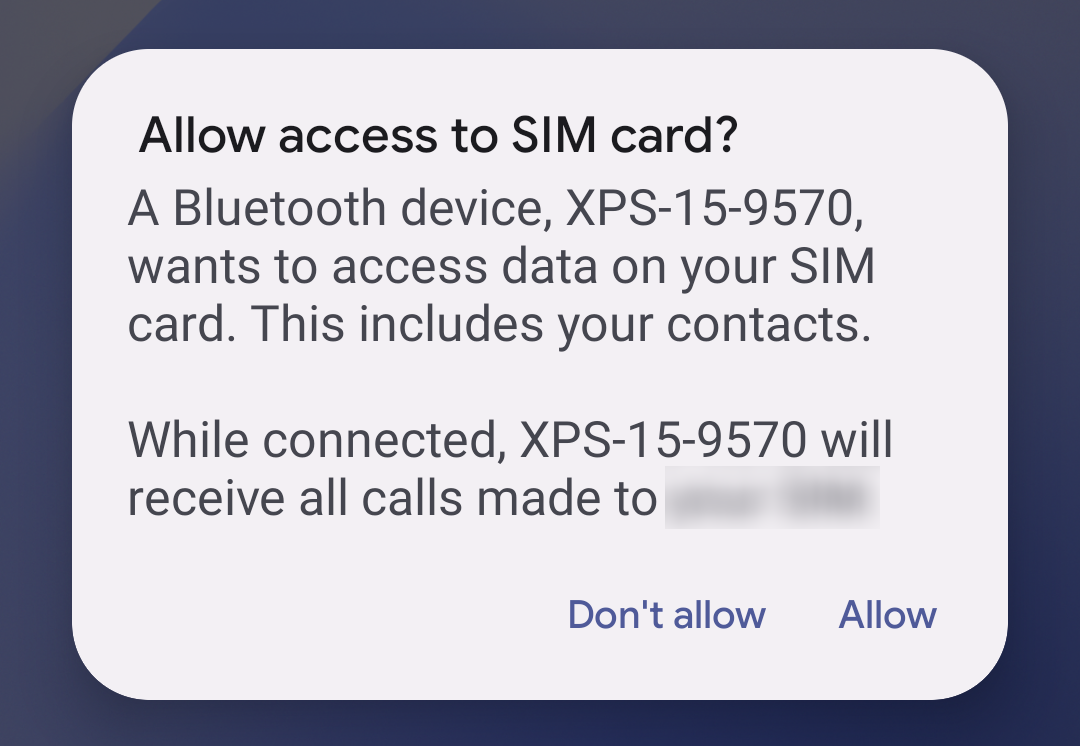}
    \caption{In addition to the SIM reader devices presented in Figure~\ref{fig:various-sim-reader}, we support eSIMs via Android's (remote) SIM Access Profile that can be accessed via Bluetooth.}
    \label{fig:android-rsap-dialog}
\end{figure}

\subsection{Attaching the Pico to Handsets and Smart Card Terminals}

\begin{figure}[b]%
    \centering%
	\includegraphics[width=.3\linewidth]{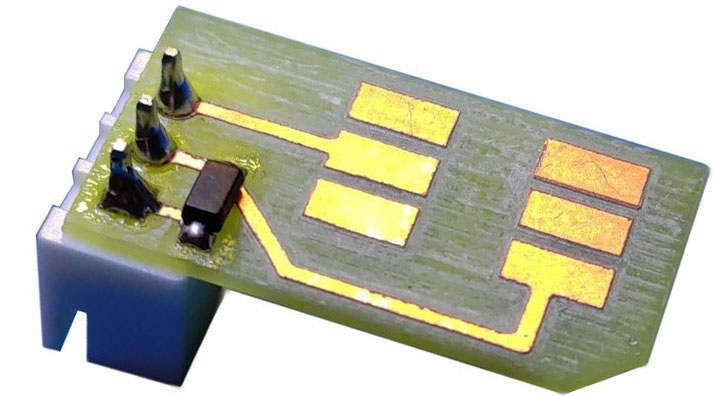}
    \caption{While improvised soldering can be used to connect to a modem or smartphone's SIM socket, we also developed a SIM adapter PCB compatible with various modem adapters.}
    \label{fig:sim-adapter}
\end{figure}

\begin{figure}[b]%
    \centering
	\includegraphics[width=.7\linewidth,trim={0 0 14cm 0},clip]{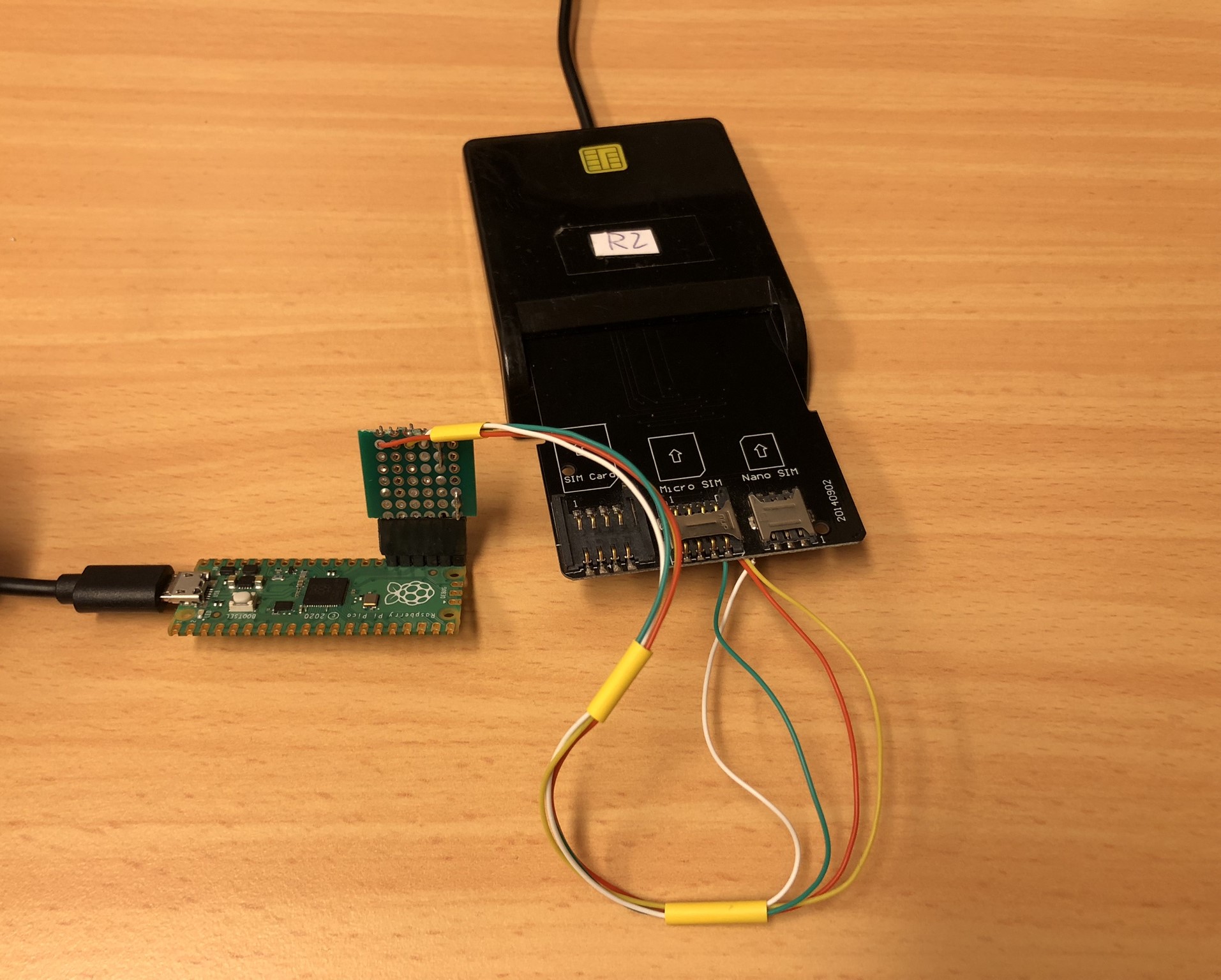}
    \caption{In addition to applications in the cellular domain, our solution supports the T=1 protocol, enabling compatibility with regular contact smart cards such as payment cards.}
    \label{fig:generic-smartcard-emulation}
\end{figure}

\clearpage



\section*{Supplementary material}

# SIMulator: SIM Tracing on a (Pico-)Budget

**Gabriel K. Gegenhuber, Philipp É. Frenzel (University of Vienna & SBA Research), Adrian Dabrwoski (FH Campus Wien)**

```
GSM SIM    67    ISO/IEC 7816-4 SELECT /EF.ICCID
GSM SIM    98    ISO/IEC 7816-4 GET RESPONSE
GSM SIM    75    ISO/IEC 7816-4 READ BINARY Offset=0
GSM SIM    67    ISO/IEC 7816-4 SELECT File EF.ELP
GSM SIM    98    ISO/IEC 7816-4 GET RESPONSE
GSM SIM    75    ISO/IEC 7816-4 READ BINARY Offset=0
GSM SIM    95    ETSI TS 102.221 TERMINAL PROFILE
```

## Problem & Motivation

▶ Inspecting, rewriting and relaying **SIM card communication** (i.e., *SIM tracing*) can be used to
  ▷ extract session keys or observe hidden SIM card communication,
  ▷ fuzz communication interfaces,
  ▷ reuse SIM cards at different locations.
▶ Professional equipment (e.g. the Osmocom SIMtrace 2) is **not available in every research lab**.
▶ The *MobileAtlas* measurement framework introduced an approach to **emulate a SIM card using basic hardware** (i.e., a serial UART interface).
  ▷ SIM card communication is tunneled over the Internet (geographic decoupling of modem and SIM).
  ▷ Allows flexible SIM card switching and reuse within all deployed measurement probes.

## Previous Work: MobileAtlas

**mobileatlas.eu**

The measurement framework can be structured into three components:
▶ **SIM providers** that allow sharing SIM card access,
▶ **measurement probes** that act as a local breakout to the cellular network, and
▶ a **management server**, that connects the prior two components and acts as command and control unit for the measurement probes.

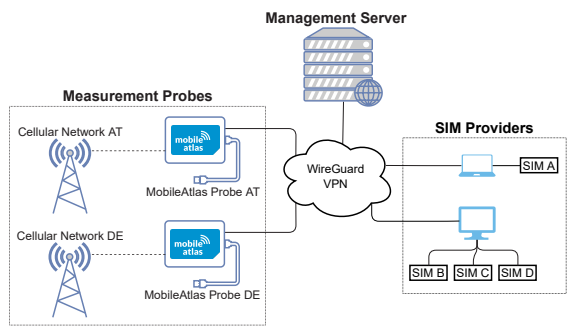

*Figure 1:* Our architecture allows every probe to use SIM cards attached to SIM providers, independent from the geographical location of probe and SIM card.

We can dynamically form a **virtual circuit** between any SIM card and measurement probe by connecting them **remotely** via a **SIM tunnel**. This boosts the **scalability** and **flexibility** of the measurement platform and allows easy measurements of **roaming** scenarios.

## SIMulator: Goals

▶ **Lower cost and entry barrier** for SIM tracing.
  ▷ Using **easily available hardware**.
▶ Leverage the existing *MobileAtlas* codebase.
  ▷ **Reduce coupling** between SIM tunnel and measurement platform.
  ▷ Add support for other smartcard types (e.g. T=0, T=1).
  ▷ Achieve broader modem (or smartphone) support.

## SIMulator: Architecture

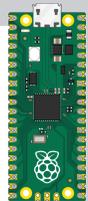

SIMulator consists of three parts:
▶ The **cellular modem** for which the SIM card is emulated.
▶ A **Raspberry Pi Pico** connected to the modem's SIM slot, providing access to the modem's SIM interface via USB.
▶ A **relaying script** running on a host system forwarding communication between the USB interface and a *MobileAtlas* SIM provider.

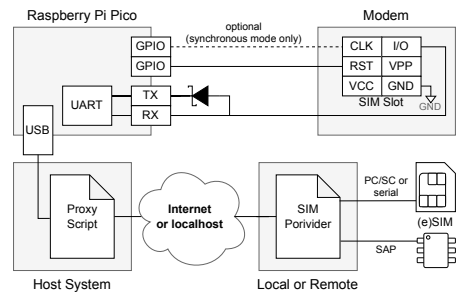

*Figure 2:* Our low-cost architecture supports full SIM tracing capabilities.

The Raspberry Pi Pico can either be used in synchronous- (UART), or in asynchronous (USART) mode:
▶ The **synchronous mode** requires an extra wire, but automatically adapts to different CLK frequencies.
▶ The **asynchronous mode** requires manually setting the CLK frequency.

## Conclusion

▶ SIMulator **considerably lowers the entry barrier** for SIM tracing.
  ▷ Only using readily available components for **less than 5 USD**.
  ▷ **Open-source** development allows reuse by other researchers.
▶ Exposing the modem's SIM interface via USB **enables novel** SIM relaying- and inspection **applications**.
  ▷ E.g., using multiple SIM tunnels and modems on one host system.
▶ Future work: **SIM virtualization** to achieve full emulation capabilities (i.e., without SIM provider).



\end{document}